# Microstreaming by ultrasound contrast microbubble between two parallel walls: shear stress and streamlines

Nima Mobadersany

Department of Mechanical and Aerospace Engineering, The George Washington University

Washington, DC 20052, USA

**Abstract**

Acoustic microstreaming has several industrial, therapeutic, and biomedical applications - Acoustic cleaning, micromixing, microfluid transport, hemolysis, sonoporation. The acoustic microstreaming due to the oscillation of ultrasound contrast microbubbles in the middle of the gap between two parallel walls is studied numerically. Axisymmetric Boundary Element Method along with theory of Nyborg is applied to study the streaming flow. Viscoelastic model with exponentially varying elasticity is used to simulate microbubble coating. The microstreaming streamlines near the plane walls and the induced shear stress is plotted. The results show the generation of a ring vortex near the walls. By increasing the gap between the walls, the radial size of the vortex ring increases while the microstreaming shear stress decreases.

## I. Introduction

An acoustic wave propagating through a medium can generate a steady streaming flow, in addition to the sinusoidal movement of the fluid particles, especially near vibrating walls (Nyborg 1953, Nyborg 1958, Riley 2001, Collis, Manasseh et al. 2010). This small-scale steady streaming flow is called microstreaming when it is caused by microbubbles. Microstreaming has several industrial applications - Acoustic cleaning and micromixing (Liu, Yang et al. 2002, Orbay, Ozcelik et al. 2017), microfluidic transport to guide solid beads and lipid vesicles in desired direction (Marmottant, Raven et al. 2006). They have also been studied extensively for biomedical applications – haemolysis (bursting of red blood cells) (Rooney 1970, Marmottant and Hilgenfeldt 2003, Pommella, Brooks et al. 2015), sonoporation (transient pore formation on cell walls under ultrasound excitation) (Fan, Kumon et al. 2014, Meng, Liu et al. 2019), drug delivery (Lentacker, De Smedt et al. 2009, Sirsi and Borden 2014, Roovers, Segers et al. 2019, Tharkar, Varanasi et al. 2019). In addition to experimental studies, there have been extensive theoretical and numerical studies elucidating the phenomena (Nyborg 1958, Wu and Du 1997, Riley 2001, Doinikov and Bouakaz 2010, Doinikov and Bouakaz 2014, Rallabandi, Wang et al. 2014, Mobadersany and Sarkar 2018, Mobadersany 2019, Inserra, Regnault et al. 2020). Lord Rayleigh (Rayleigh 1945) was the first who derived the first theoretical explanation of the acoustic streaming phenomena observed in Kundt's tube due to small oscillatory motion. This was followed in the 1950s by a series of pioneering studies (Westervelt 1953, Nyborg 1958, Lighthill 1978). Specifically, Nyborg provided a generalized perturbative analysis of the streaming velocity near a boundary and offered expressions for the induced stress, which have been widely used in the literature. Rooney (Rooney 1970) used Nyborg's theory and calculated maximum shear stress induced by pulsating bubbles resting on a wall for hemolysis of red blood cells. Levin and Bjorno(Lewin and Bjo/rno/ 1982) studied the maximum shear stress due to the gases microbubbles on biological cells using Nyborg's theory. Wu, et al. (Wu, Ross et al. 2002) used a modified Rayleigh–Plesset equation with a shell model due to (de Jong, Cornet et al. 1994) and the same expression to show that it can predict a shear stress (~12 Pa) in the streaming field of an Optison contrast microbubble at ~0.1 MPa and 1 or 2 MHz excitation, which is sufficient for reparable sonoporation in a living cell. Forbes and O'Brien (Forbes and O'Brien 2012) used a similar methodology using the Marmottant model (Marmottant, van der Meer et al. 2005) for the encapsulation to predict an increase of sonoporation. Krasovitski and Kimmel (Krasovitski and Kimmel 2004) were the first applied Nyborg theory for the pulsating bubble detached from the wall. They used axisymmetric boundary element method along with Nyborg expression and calculated shear stress on the wall induced by encapsulated microbubbles. The encapsulation is assumed as a thin elastic non-compressible shell. Considering the linear oscillation amplitude of spherically pulsating microbubble, Doinikov and Bouakaz (Doinikov and Bouakaz 2010) applied Nyborg's theory and calculated the microstreaming shear stress on the plane wall analytically assuming the bubble is detached from the wall. Several researches have conducted theoretical study on the microstreaming streamlines inside and outside pulsating microbubbles. (Wu and Du 1997) computed the microstreaming flow field inside and outside an isolated microbubble oscillating in the field of a plane ultrasound wave by accounting for the monopole volume pulsation and the dipole translation motion in the spherical geometry. The results later were further generalized by removing restrictions on bubble size relative to the wavelength and considering viscous effects in the whole domain (Doinikov and Bouakaz 2010). Detailed analytical theories have also been developed by Doinikov to show that the microstreaming near an oscillating bubble increases considerably due to the presence of a distant rigid wall or in the presence of a second bubble (Doinikov and Bouakaz 2014), especially when they are driven at their resonance frequency. Inserre, et al. (Inserra, Regnault et al. 2020) developed an analytical theory and showed the microstreaming around non-spherical oscillation of a gas bubbles. However there have been very few studies for the streamlines of the microstreaming flow field near a wall due to detached microbubble from the wall. In our earlier study, closed streamlines of microstreaming flow near a plane wall due to pulsation of free and coated microbubble using analytical expression were shown for the first time while keeping the bubble in further distance from the wall in order to assume spherical oscillation (Mobadersany and Sarkar 2019). However, there are several occasions where the microstreaming is of high interest in a channel for microfluidic transport or in a blood vessel for sonoporation which makes it important to understand the microstreaming due to microbubble confined between two walls. Therefore, in the present work, a numerical study is conducted on microstreaming flow due to the pulsation of coated microbubble in the middle of the gap between two plane walls. Axisymmetric boundary element method is used for the numerical study. Local irrotational velocity near the plane walls were computed using the boundary element method. Then, revisited theory of Nyborg was used for computing microstreaming streamlines and the induced shear stress on the wall due to ultrasound contrast microbubbles. The microbubble was assumed as Sonazoid; high density gas core ultrasound contrast agent encapsulated with a monolayer of lipid. The shell of the contrast agent is modeled as a viscoelastic interface.

## II. Numerical method

### A. Governing equations

Theoretical analysis of microstreaming solves the governing Naiver Stokes equations by a perturbative method (Nyborg 1953, Nyborg 1958):

$$\mathbf{u} = \mathbf{u}^{(1)} + \mathbf{u}^{(2)} + ...$$
$$p = p^{(1)} + p^{(2)} + ... \tag{1}$$

The first order approximation solves the linearized Naiver Stokes equation neglecting the nonlinear advection terms and obtains a sinusoidal velocity. At the second order, the convective nonlinear term, quadratic product of $\mathbf{u}^{(1)}$, appears as a forcing term, with the equation upon averaging gives the governing equation of the streaming motion(2).

$$\mu \nabla^2 <\mathbf{u}^{(2)}>_t - \nabla <p^{(2)}>_t = \mathbf{F} = \rho <\mathbf{u}^{(1)} \cdot \nabla \mathbf{u}^{(1)}>_t, \tag{2}$$

with $\rho$ and $\mu$ being the fluid density and viscosity, $<>_t$ is the average over the time period of the oscillating excitation, $<\mathbf{u}^{(2)}>_t$ is the streaming velocity vector (second order approximation of the velocity), $\mathbf{u}^{(1)}$ is the sinusoidal velocity vector (first order approximation of the velocity).

Note that the resulting streaming motion has been named Rayleigh-Nyborg-Westervelt (RNW) streaming by (Lighthill 1978) in contrast to Stuart streaming (Stuart 1966). The underlying approximation is only valid when bubble streaming Reynolds number $\text{Re}_{bs}$ is small. In anticipation of the streaming velocity $\sim U^2/\omega R_0$ and the length scale $\delta = \sqrt{2\nu/\omega}$ ($\nu = \mu/\rho$ kinematic viscosity), one obtains (Davidson and Riley 1971, Marmottant and Hilgenfeldt 2003) $\text{Re}_{bs} = \varepsilon^2 \left(2\omega R_0^2/\nu\right)^{1/2} = \varepsilon^2 (2\mathcal{R}_{e_b})^{1/2}$, with $\mathcal{R}_{e_b} = \omega R_0^2/\nu$ being the Reynolds number of the bubble motion. Solving(2), Nyborg (Nyborg 1958) found the radial component of the streaming velocity for a flow over a plane rigid wall as the function of local irrotational velocity. The streaming velocity $<u^{(2)}>_t$ for an axisymmetric flow in radial direction near the plane wall is as follows (for more details (Mobadersany and Sarkar 2019)

$$<u^{(2)}>_t = \frac{1}{\omega}\left(\frac{1}{2}\frac{\partial u_\varphi^2}{\partial r}(u_\alpha - u_\beta) - \frac{u_\varphi^2}{r}u_\beta\right),$$

$$u_\alpha = \frac{1}{4}e^{-2\beta z} + e^{-\beta z}\sin\beta z - \frac{1}{4}, \tag{3}$$

$$u_\beta = \frac{1}{2}\beta z e^{-\beta z}(\cos\beta z - \sin\beta z) - e^{-\beta z}\left(\sin\beta z + \frac{1}{2}\cos\beta z\right) + \frac{1}{2}.$$

$\beta^{-1} = \delta$ is the acoustic boundary layer, $\omega$ is the excitation angular frequency, $r$ is the radial distance, $z$ is the vertical distance from the plane wall and $u_\varphi$ is the amplitude of the local radial irrotational velocity $u^{(\varphi)} = u_\varphi \cos \omega t$. The vertical component of streaming velocity $<w^{(2)}>_t$ is obtained by using equation of mass conservation and considering the no slip condition on the plane wall.

$$\frac{\partial \left(r <w^{(2)}>_t\right)}{\partial z} + \frac{1}{r}\frac{\partial \left(r <u^{(2)}>\right)}{\partial r} = 0. \tag{4}$$

The acoustic streaming velocity field is therefore known in terms of the outer irrotational velocity field. In this study, Boundary Element Method (BEM) for potential flow has been used to calculate the local irrotational fluid velocity field near the plane wall. Knowing the irrotational velocity, one can compute the shear stress on the wall (Nyborg 1958):

$$\tau_{wall} = \mu \left.\frac{\partial <u^{(2)}>_t}{\partial z}\right|_{z=0} = \frac{\rho_0}{4\beta} u_\varphi \left.\frac{\partial u_\varphi}{\partial r}\right|_{z=0}. \tag{5}$$

**B. Computing irrotational velocity using Boundary Element Method (BEM)**

To find the local irrotational velocity, Laplace equation has been solved for the flow around the microbubble resulted in the following Green's integral formula.

$$c(p)\varphi_i + \sum_{j=1}^{N} \int_{S_j} \varphi_j \frac{\partial}{\partial n}\left(\frac{1}{|p_i - q_j|}\right) ds = \sum_{j=1}^{N} \int_{S_j} \frac{\partial \varphi_j}{\partial n}\left(\frac{1}{|p_i - q_j|}\right) ds. \tag{6}$$

In equation(6), $S$ is the boundary of the liquid domain which includes the wall of the coated microbubble and the interfaces of the liquid domain with two parallel plane walls; $\varphi$ is the velocity potential and $\partial\varphi/\partial n$ is the normal velocity; $q$ is any point on the fluid boundary (microbubble surface and plane walls); $P$ is any point in the fluid domain, or on its boundary. $c(p)$ is a coefficient dependent on the location of point $P$; $2\pi$ for points on the coated microbubble and $4\pi$ for points inside the fluid domain. $N$ is the number of elements on the boundary. It should be noted that, the discretization of the upper and lower parallel walls is extended up to physical infinity (Shervani-Tabar and Mobadersany 2011).

Figure 1 shows the schematic of the discretized geometry. The wall of the coated microbubble is discretized into cubic spline elements and the parallel walls are discretized into linear elements. Collocation points are located at the middle of each element and velocity potential and the normal velocity are assumed to be constant on each element.

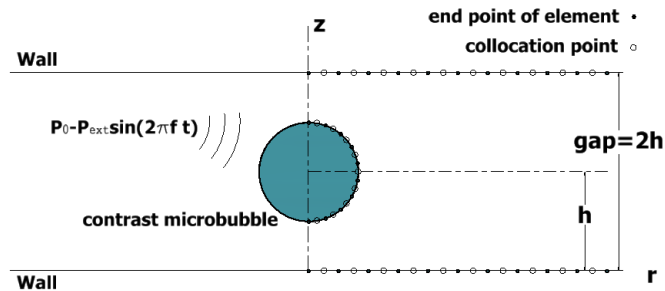

**Figure 1 Schematic of the problem and discretized geometry**

The normal vector on the boundary of the fluid domain is indicated by $n$ and is directed outward the fluid domain. The problem is assumed axisymmetric, and therefore $\varphi$ and $\partial\varphi/\partial n$ are independent of azimuth angle. $z$ is the vertical axis, and $r$ is the radial axis of the axisymmetric geometry. $2h$ is the gap size; the distance between the two parallel walls. Ultrasound contrast microbubble is excited in the middle of the gap located at distance $h$ from each plane wall. Equation (6) gives a set of linear equations for evaluating $\partial\varphi/\partial n$ on each segment on the microbubble wall and $\varphi$ on each segment of the plane walls. Note that velocity potential is zero on the microbubble wall at initial time. Normal velocity is also zero on the plane walls. The evaluation of the elements of the integrands is calculated numerically using Gauss-Legendre quadrature, unless the collocation point is within the segment and therefore the integrand is singular. The singular integrals have been treated especially (for more details see (Taib 1985)). By having distribution of the velocity potential on the microbubble wall, tangential velocity on the microbubble wall is obtained by differentiation of the velocity potential along the wall of the microbubble. By having the velocity of each element on the microbubble wall, the shape of the coated microbubble at the next time step is obtained using the second order Runge-Kutta scheme. Similarly, the velocity potential is being updated using the second order Runge-Kutta scheme (Shervani-Tabar and Mobadersany 2013). As discussed above, to solve the set of linear equations given in (6), the velocity potential on the microbubble wall should be known at each time step. To find the velocity potential in the next time step, the unsteady Bernoulli equation is solved on each element.

$$\rho\left(\frac{D\varphi}{Dt} - \frac{1}{2}|\nabla\varphi|^2\right) = P_\infty - P_b. \tag{7}$$

$P_b$ is the pressure of the fluid on the bubble wall. $P_\infty = P_0 - P_{ex}\sin(2\pi f\,t)$ is the pressure in the far field which includes the static ambient pressure $P_0$ (here atmospheric pressure) and ultrasound excitation pressure. $P_{ex}$ and $f_{ex}$ are the amplitude of the ultrasound excitation pressure and frequency. The problem under investigation is non-dimensionalized by employing the initial radius of the ultrasound contrast microbubble $R_0$, the liquid density $\rho$, and the excitation angular frequency $\omega$. Normalized form of (7) is:

$$\frac{D\phi^*}{Dt^*} = \frac{1}{2}|\nabla^*\phi^*|^2 + \frac{(P_\infty - P_b)}{\rho R_0^2 \omega^2}. \tag{8}$$

$\varphi^* = \varphi/(\omega R_0^2)$ is the non-dimensional velocity potential, $t^* = t\omega$ is the non-dimensional time and $\nabla^* = R_0\nabla$ is the non-dimensional gradient. The non-dimensional velocity potential is computed using the following non-dimensional variable time step.

$$\Delta t^* = \frac{\Delta\varphi^*}{\max\left(\frac{P_\infty - P_b}{\rho R_0^2 \omega^2} + \frac{1}{2}|\nabla^*\varphi^*|^2\right)}. \tag{9}$$

$\Delta\varphi^*$ is some constant that controls the maximum increment of the velocity potential on the microbubble surface between two successive time steps. Once the instant velocity potential and normal velocity on the boundary of the fluid domain (microbubble surface and plane walls) are known, the velocity potential at any point inside the liquid domain can be found by using Green's integral formula in equation (6).

By knowing the velocity potential in the liquid near the plane wall at each time step, the instant irotational local velocity $u^{(\varphi)} = u_\varphi \cos \omega t$ is obtained by employing a finite difference scheme. The amplitude of the non-dimensional local irrotational velocity can be found over an oscillation cycle of ultrasound contrast microbubble in the following:

$$u_\varphi^* = \frac{\left(\nabla^* \varphi^*\right)_{\max} - \left(\nabla^* \varphi^*\right)_{\min}}{2}. \tag{10}$$

Note that the oscillation cycle has been chosen in the steady state region when the transient oscillation of microbubble has already been subsided.

## C. Modelling the encapsulation of ultrasound contrast microbubble

The pressure inside and outside the microbubble are related by the normal stress balance. The effect of the coating was considered in the normal stress balance. The encapsulation of the microbubble is assumed as a viscoelastic interface (Chatterjee and Sarkar 2003) having both elastic wall tension and dilatational viscosity (Scriven 1960).

$$P_b = P_g - \left(\sigma_{\text{eff}} + \kappa^s \left(\nabla_s \cdot V\right)\right)\left(\nabla_s \cdot n\right). \tag{11}$$

In equation(11), $P_g = P_{g_0} \left(V_0/V\right)^\kappa$ is the gas pressure inside the microbubble when its volume is $V$. The gas inside the microbubble is assumed to be an ideal gas following polytropic behavior. $P_{g_0}$ is the initial gas pressure inside the microbubble when it is in its initial volume $V_0$, and $\kappa$ is the polytropic constant. $\sigma_{\text{eff}}$ and $\kappa^s$ are the effective surface tension and dilatation viscosity of the ultrasound contrast microbubble due to the encapsulation, $\nabla_s \cdot V$ is the surface divergence of velocity, and $\nabla_s \cdot n$ is the curvature on each element of the microbubble.

$$\begin{aligned}\nabla_s \cdot n = &\left(\frac{dz}{d\xi}\frac{d^2 r}{d\xi^2} - \frac{dr}{d\xi}\frac{d^2 z}{d\xi^2}\right) \bigg/ \left(\left(\frac{dr}{d\xi}\right)^2 + \left(\frac{dz}{d\xi}\right)^2\right)^{3/2} \\ &- \left(\frac{dz}{d\xi}\right) \bigg/ \left(r(\xi)\left(\left(\frac{dr}{d\xi}\right)^2 + \left(\frac{dz}{d\xi}\right)^2\right)^{1/2}\right).\end{aligned} \tag{12}$$

$\xi$ is the arc length parameter describing $r$ and $z$. The effective surface tension $\sigma_{\text{eff}}$ describes the interfacial rheology of the encapsulation. Several models have been proposed to study the interfacial rheology, i.e. $\sigma_{\text{eff}}$. In the present study, a viscoelastic model with exponentially varying elasticity (EEM) (Paul, Katiyar et al. 2010) is used to simulate the effective wall tension due to the presence of microbubble coating.

$$\sigma_{eff}(R) = \sigma_0 + \beta E_0^s \exp(-\alpha^s \beta),$$

$$\beta = (\frac{R^2}{R_E^2} - 1), \quad R_E = R_0[1 + (\frac{1 - \sqrt{1 + 4\sigma_0 \alpha^s / E_0^s}}{2\alpha^s})]^{-1/2}. \tag{13}$$

$\sigma_{eff}(R)$ is the effective interfacial tension which is the function of instant radius of the microbubble. $E_0^s$ is elasticity constant and $\beta$ is the area fraction. $R_E$ is the equilibrium radius of the coated microbubble where elastic stress is zero. Equation (13) have been developed for the encapsulation rheology of ultrasound contrast microbubbles oscillating spherically. But the oscillation of the microbubble near the walls may not be purely spherical, therefore each wall element on the microbubble may have different area fraction which results in different effective surface tension for each segment. Therefore, one can modify the area fraction in equation (13) as:

$$\begin{cases} \beta = (\frac{A}{A_E} - 1), \\ A_E = A_0[1 + (\frac{1 - \sqrt{1 + 4\sigma_0 \alpha^s / E_0^s}}{2\alpha^s})]^{-1}. \end{cases} \tag{14}$$

$A$, $A_0$ and $A_E$ are the surface area, initial surface area and equilibrium surface area of each element on the ultrasound contrast microbubble. In this study, the properties of Sonazoid are used as the desired ultrasound contrast agent. The constant surface tension, constant elasticity and constant elastic coefficients in equation (13) for Sonazoid contrast agent are $\sigma_0 = 0.019 \, N/m$, $E_0^s = 0.55 \, N/m$, $\alpha^s = 1.5$ and $\kappa = 1.07$ respectively (Katiyar and Sarkar 2012).

### III. Results and discussion

In this study, the main interest is in microbubble applications for ultrasound imaging and drug delivery. Therefore, Sonazoid contrast agent is used as the ultrasound contrast microbubble in this study. The initial size of the contrast microbubble is considered $R_0 = 1.6 \mu m$; the average radius size of Sonazoid contrast agent. The microbubble is excited with ultrasound wave of 5MHz frequency which is within typical frequency range for biomedical applications. The excitation pressure $P_{ex}$ in this study is low enough to make the microbubble oscillate linearly. The study has been carried for ultrasound contrast microbubble oscillating in the middle of the gap between two parallel plane walls, so that the plane wall exert equal secondary radiation force on the microbubble causing it to stay in the middle of the gap and oscillate steadily.

#### A. Validation of the numerical method
To validate the numerical method, the oscillation of contrast microbubble has been compared with modified Rayleigh-Plesset type equation (modified RP equation) (15) which can only capture spherical microbubble dynamics. The numerical method has been validated against the spherical oscillation of the coated microbubble near one plane wall since there were not enough material in the literature to compare the microbubble oscillation between two parallel walls.

$$\rho\left(R\ddot{R}+\frac{3}{2}\dot{R}^2\right)=P_b-P_\infty,$$

$$P_b = P_{g_0}\left(\frac{R_0}{R}\right)^{3k} - 4\kappa^s\frac{\dot{R}}{R^2} - \frac{2\sigma_{eff}(R)}{R} - P_s(h,t), \qquad (15)$$

$$P_s(h,t) = \rho\frac{R}{2h}\left(R\ddot{R}+2\dot{R}^2\right).$$

$\dot{R}$ and $R$ are the velocity and instant radius of the contrast microbubble. In modified RP equation, a virtual bubble has been considered to replace the plane wall to satisfy the impermeability condition. Note that the effect of the plane wall is considered as a scattered pressure $P_s(h,t)$ by the the virtual bubble pulsating at the distance $2h$ from the contrast microbubble. To have a valid comparison between the results from Boundary Element Method (BEM) and the modified RP equation (15), the microbubble was excited at a distance far from the plane wall in order to maintain spherical shape while pulsating.

Figure 2 shows the radial oscillation of the ultrasound contrast microbubble with respect to non-dimensional time using BEM and modified RP equation for comparison. The microbubble is initially located at $h = 3.75R_0$; far enough from the plane wall so that the bubble maintains its spherical shape. The comparison of radial oscillation of coated microbubble near a wall shows that numerical results from BEM are in good agreement with the numerical results from modified RP equation.

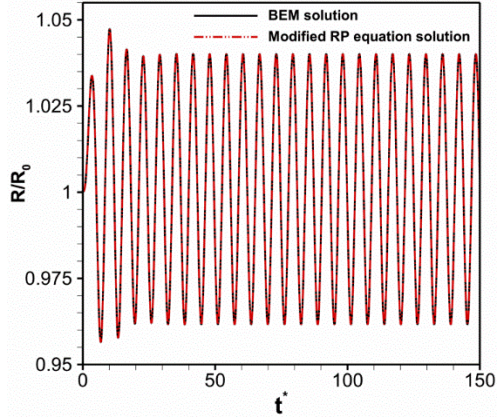

**Figure 2 Radial oscillation of contrast microbubble in the presence of one wall with respect to non-dimensional time using BEM and modified RP equation when $P_{ex} = 60KPa$, $f = 5MHz$, $h = 3.75R_0$**

Note that equations (5) and (15) can be linearized assuming the microbubble is pulsating with a small amplitude $R = R_0(1+x)$ where $x = \xi_m \sin(\omega t + \phi)$ to give the analytical expression of the microstreaming shear stress on the wall (for more details (Mobadersany and Sarkar 2019)).

$$\frac{\tau_{wall}}{P_0} = \frac{2\xi_m^2 (r/R_0)}{\mathcal{E}_{u_b}(2\mathcal{R}_{e_b})^{1/2}(h/R_0)^5} \frac{(1-2r^2/h^2)}{(1+r^2/h^2)^4} \tag{16}$$

$\mathcal{R}_{e_b} = \rho R_0^2 \omega / \mu, \mathcal{E}_{u_b} = P_0 / \rho R_0^2 \omega^2$ are the characteristic Reynolds ($R_0\omega$ is taken as the velocity scale) and Euler numbers. Equation (16) is the analytical expression of the microstreaming shear stress for the microbubble oscillating near one plane wall. Note that equation (16) is derived from equation (5) using perturbation theory assuming the microbubble is pulsation linearly. Figure 3 shows the microstreaming shear stress on the plane wall induced by pulsation of contrast microbubble for the conditions mentioned in figure 2 where the microbubble is located far enough from the plane wall so that it maintains its spherical shape. The microstreaming shear stress on the wall has been plotted using BEM and the analytical expression for spherically pulsating microbubble near one wall (equation 16). As shown in figure 4, the shear stress obtained from BEM is in good agreement with the shear stress from analytical solution (equation 16).

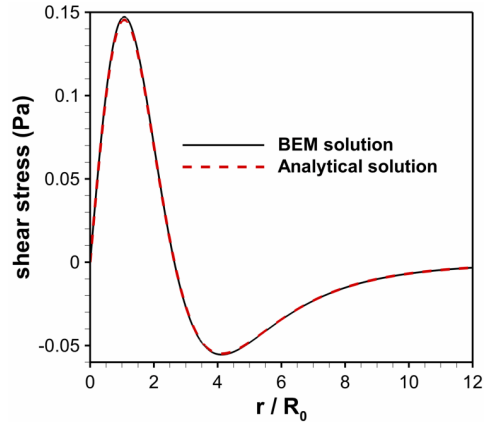

**Figure 3 Microstreaming shear stress due to oscillation of contrast microbubble in the presence of one wall when $P_{ex} = 60KPa$, $f = 5MHz$, $h = 3.75R_0$**

### B. Numerical results

To observe the behavior of the contrast microbubble confined between two parallel walls, figure 4(a-b) show the radial oscillation and microstreaming shear stress of the contrast microbubble oscillating in the middle of the gap between two parallel plane walls. The contrast microbubble is located at equal distance $h = 3.75R_0$ from each wall ($gap = 7.5R_0$). The oscillation of contrast microbubble and microstreaming shear stress are compared with the conditions mentioned in figure 3 where the contrast microbubble is pulsating near one plane wall at $h = 3.75R_0$.

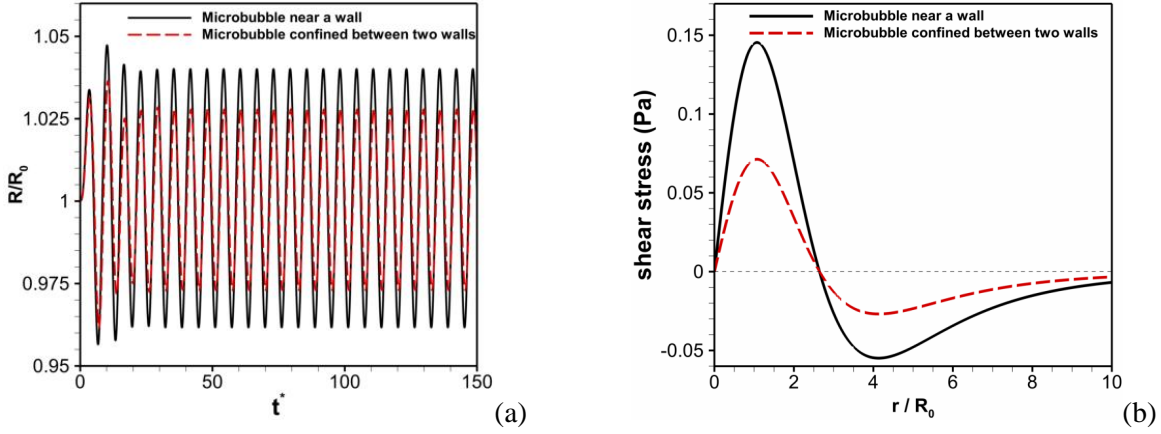

**Figure 4** Comparison of contrast microbubble behavior in the presence of one wall and between two walls when $P_{ex} = 60 KPa$, $f = 5MHz$, $h = 3.75R_0$ (a) radial oscillation (b) corresponding microstreaming shear stress on the wall

As shown in figure 4(a), the contrast microbubble has smaller oscillation amplitude when it is located between two walls. This results in lower microstreaming shear stress on the walls as seen in figure 4(b). As it is observed, the radial position where the sign of the shear stress changes (becomes zero) is the same for both case; where the microbubble is between two parallel walls and when the microbubble is oscillating near one wall. Not that the shear stress on one of the parallel walls have been shown in figure 4(b) since the microbubble is oscillating in the middle of the gap which gives rise to the same shear stress on each parallel wall.

**B.1. Effect of gap between the parallel walls**

Figure 5(a) shows the streamlines of the microstreaming flow near the parallel plane walls induced by ultrasound contrast microbubble located at $h = 2R_0$ from each wall in the middle of the gap and excited at 0.18MPa. As it is shown, a ring vortex is generated near each plane wall. Figure 5(b) shows the induced microstreaming shear stress on the plane walls. As seen in figure 5(b), shear stress is positive at radial distances below 1.4R0. The radial length of the vortex can be estimated by the location where the direction of the shear stress (equation (5)) on the wall is changing. This location corresponds to the radial distance on the plane wall where the amplitude $u_\varphi$ of the potential velocity $u^\varphi = u_\varphi \cos(\omega t)$ becomes maximum (see figure 6). As it has already been shown in figure 4, the radial position of the point where the direction of the shear stress changes remains the same for the microbubble oscillating near one wall and for the microbubble oscillating between two parallel walls while keeping the same vertical distance from the wall. Therefore, the radial length of the vortex will be the same for the two cases. We have previously developed a theory (Mobadersany and Sarkar 2019) showing the microstreaming radial vortex length for a microbubble oscillating near a plane wall.

$$\frac{L_{vortex}}{R_0} = \frac{h}{\sqrt{2}R_0} \tag{17}$$

where $L_{vortex}$ is the radial length of the vortex ring which seems to be valid for the microbubble oscillating between two parallel walls. This shows that the vortex length is the function of $h$, the microbubble separation from the plane walls. In contrast to the vortex length, its vertical extent $d_{vortex}$ remains unchanged with varying $h$ since it is of the order $\sim \delta = \sqrt{2\nu/\omega}$, the boundary layer thickness (for more details (Mobadersany and Sarkar 2019 )).

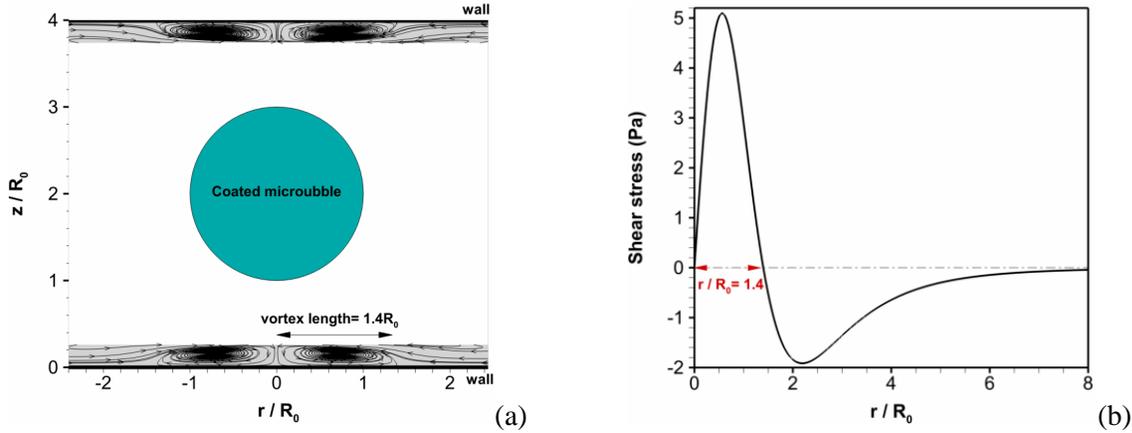

**Figure 5 (a) Microstreaming streamlines near plane walls, (b) microstreaming shear stress on each wall when $P_{ex} = 180 KPa$, $f = 5 MHz$, $h = 2R_0$ ($gap = 4R_0$)**

The change of the sign of the microstreaming shear stress is due to the non-monotonic behavior of the velocity potential on the plane wall. Figure 6 shows the amplitude of non-dimensional radial potential velocity $u_\varphi^*$ on the parallel plane walls due to the pulsation of contrast microbubble for the conditions mentioned in figure 4. As it is shown, the velocity potential has a maximum value at $r = 1.4R_0$. This causes a change in the direction of the shear stress at $r = 1.4R_0$, and the force that drives microstreaming. It is the change in the direction of the driving external force which pushes the fluid upward (as the flow cannot go downward because of the wall) and creates circulatory flows near the plane wall.

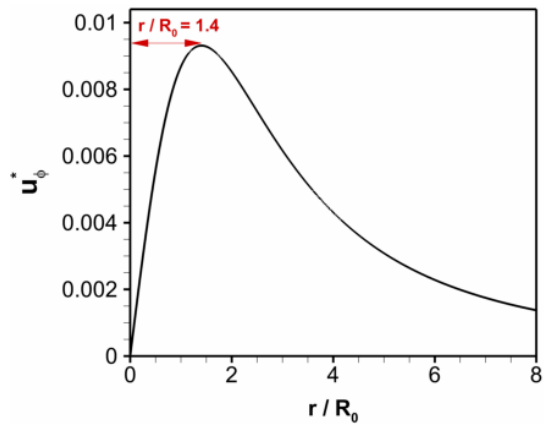

**Figure 6 Amplitude of normalized potential velocity on the plane wall due to oscillation of contrast microbubble when $P_{ex} = 180 KPa$, $f = 5 MHz$, $h = 2R_0$ ($gap = 4R_0$)**

Figure 7 shows the microstreaming due to contrast microbubble when the parallel walls are closer to each other (the gap between the walls is smaller).

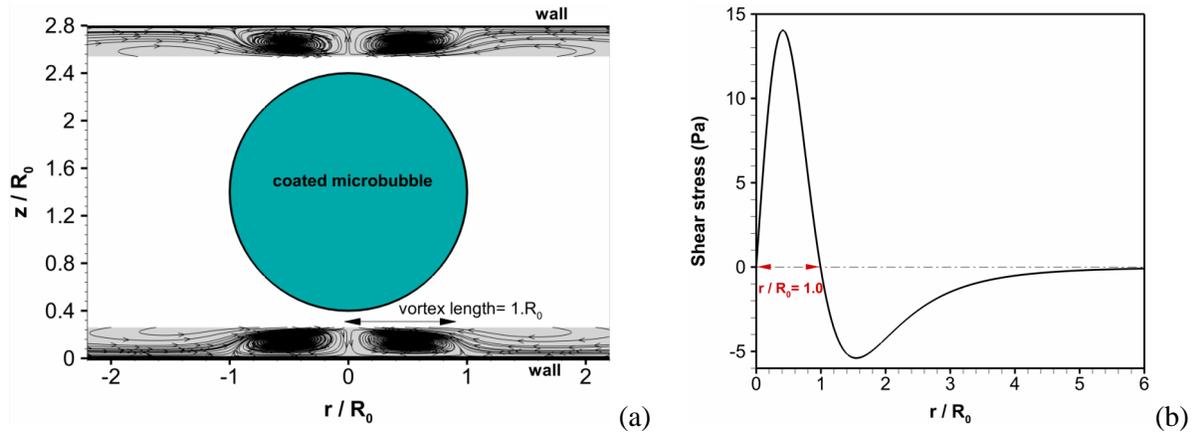

**Figure 7 (a) Microstreaming streamlines near plane walls, (b) microstreaming shear stress on each wall when** $P_{ex} = 180 KPa, \ f = 5MHz, \ h = 1.4R_0 \ (gap = 2.8R_0)$

As it is seen in figure 7(a), the radial size of the ring vortex becomes $1.0R_0$. Therefore, the vortex radial size of the ring vortex decreases as the microbubble is excited at closer distances to the plane walls. This is consistent with equation (16) showing the vortex radial length is decreasing to $1.0R_0$ with decreasing the gap size. On the contrary, the microstreaming shear stress on the walls is increasing.

**B.2 Effect of ultrasound excitation frequency**

As shown in Figures 5(b) and 7(b), the ultrasound contrast microbubble induces positive (outward radial direction) and negative microstreaming shear stress (inward radial direction) on the walls. The shear stress has one peak at the region when the wall experiences positive shear stress and another peak at the region when the wall experiences negative shear stress. The radial oscillation amplitude of ultrasound contrast microbubble and the peak of induced shear stress value (positive shear stress value or negative shear stress value) at different excitation frequencies is shown in Figure 8 when the gap between the walls is $4R_0$. Note that shear stress peak value (positive or negative) in figure 8 is normalized by the largest peak value of shear stress (positive or negative) within the studied frequency range. Note that normalized shear stress peak value is the same for both negative and positive shear stress peak. As it is observed, the contrast agent has the maximum oscillation amplitude when it is excited at 2.80MHz indicative of the resonance frequency of the contrast agent when the gap between the walls is $4R_0$. The shear stress has the maximum value at 2.85MHz which is slightly higher than the resonance frequency. Note that the maximum shear stress does not occur exactly at resonant frequency. As seen in equation (16), microstreaming shear stress is changing directly by both excitation frequency and oscillation amplitude of microbubble. Therefore, even though the microbubble has maximum oscillation amplitude at resonance frequency, the maximum shear stress occurs at the excitation frequency slightly higher than resonance. Figure 9 shows the radial oscillation amplitude of ultrasound contrast microbubble and the peak of induced shear stress value (positive shear stress value or negative shear stress value value) at different excitation frequencies when the gap between the walls is $2.8R_0$. Peak value of shear stress (positive or negative) in figure 9 is normalized by the largest peak value of shear stress (positive or negative) within the studied frequency range. Again, as it is observed in figure 9, the maximum bubble oscillation occurs at 2.45MHz while the maximum shear stress value occurs at 2.5MHz. It is worth mentioning that the presence of walls reduces the resonance frequency of ultrasound contrast microbubble. Based on the properties and encapsulation model used in this study, in the absence of any wall, the ultrasound contrast agent has natural frequency (undamped resonance frequency) of:

$$f_n = \left( \sqrt{\left( 3\kappa p_0 + 2E_0^s/R_0 \left( \sqrt{1+4\sigma_0 \alpha^s/E_0^s}/\alpha^s \right)\left( 1+2\alpha^s - \sqrt{1+4\sigma_0 \alpha^s/E_0^s} \right) \right)/\rho} \right) \Big/ 2\pi R_0 = 4.2 MHz$$

and resonance frequency of $f_r = f_n\sqrt{1-0.5\delta_t^2} = 3.9 MHz$,

where $\delta_t = 4\mu/\left(2\pi\rho f_n R_0^2\right) + 4\kappa^s/\left(2\pi\rho f_n R_0^3\right)$ is the total damping due to the surrounding liquid and shell dilatational viscosity. The resonance frequency of contrast agent increases when the gap between the walls increase.

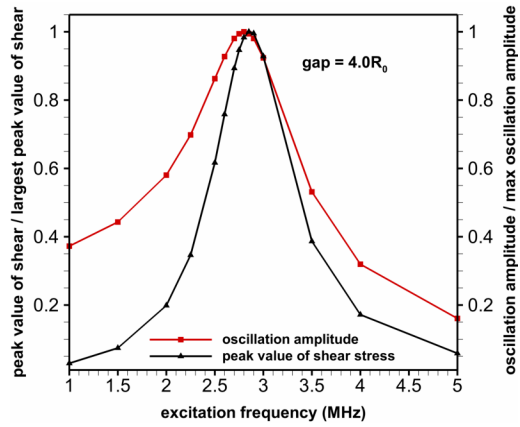

**Figure 8 Oscillation amplitude of contrast microbubble and normalized maximum peak value of shear stress with respect to excitation frequency when** $gap = 4R_0 \ (h=2R_0)$

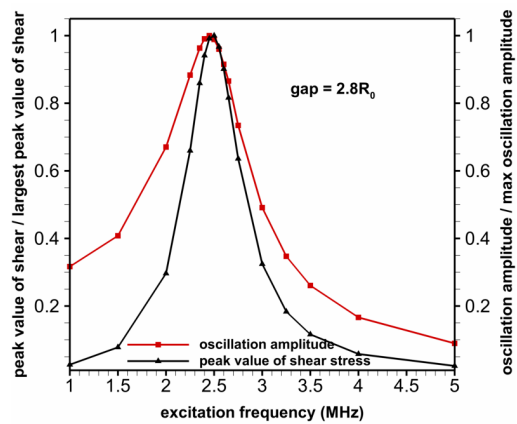

**Figure 9 Oscillation amplitude of contrast microbubble and normalized peak value of shear stress with respect to excitation frequency when** $gap = 2.8R_0 \ (h=1.4R_0)$

## IV. Conclusion

In this study, the microstreaming flow due to the pulsation of ultrasound contrast microbubbles in the middle of the gap between two plane walls have been studied numerically along with the induced microstreaming shear stress. The boundary element method has been used to calculate the local potential velocity component of fluid particles in microstreaming velocity field. The encapsulation of the microbubble is considered as a viscoelastic interface. Exponentially elasticity varying model (EEM) has been used to simulate the rheology of the microbubble encapsulation. A ring vortex has been generated near each plane wall with the similar vortex length that observed for a contrast microbubble pulsating in the presence of one plane wall. But the microstreaming shear stress is lower when the microbubble is oscillating between two walls. It is shown that decreasing the gap between the plane walls results in the increase of the shear stress and decrease of the radial distance of the ring vortex. It is also shown that the contrast microbubble has higher resonance frequency when the distance between the walls (gap) is larger. The maximum shear stress occurs at excitation frequency slightly higher than the resonance.

## References

1. Chatterjee, D. and K. Sarkar (2003). "A Newtonian rheological model for the interface of microbubble contrast agents." Ultrasound in Medicine and Biology **29**(12): 1749-1757.
2. Collis, J., R. Manasseh, P. Liovic, P. Tho, A. Ooi, K. Petkovic-Duran and Y. G. Zhu (2010). "Cavitation microstreaming and stress fields created by microbubbles." Ultrasonics **50**(2): 273-279.
3. Davidson, B. J. and N. Riley (1971). "Cavitation Microstreaming." Journal of Sound and Vibration **15**(2): 217-233.
4. de Jong, N., R. Cornet and C. T. Lancée (1994). "Higher harmonics of vibrating gas-filled microspheres. Part one: simulations." Ultrasonics **32**(6): 447-453.
5. Doinikov, A. A. and A. Bouakaz (2010). "Acoustic microstreaming around a gas bubble." Journal of the Acoustical Society of America **127**(2): 703-709.
6. Doinikov, A. A. and A. Bouakaz (2010). "Theoretical investigation of shear stress generated by a contrast microbubble on the cell membrane as a mechanism for sonoporation." Journal of the Acoustical Society of America **128**(1): 11-19.
7. Doinikov, A. A. and A. Bouakaz (2014). "Effect of a distant rigid wall on microstreaming generated by an acoustically driven gas bubble." Journal of Fluid Mechanics **742**: 425-445.
8. Fan, Z., R. E. Kumon and C. X. Deng (2014). "Mechanisms of microbubble-facilitated sonoporation for drug and gene delivery." Therapeutic delivery **5**(4): 467-486.
9. Forbes, M. M. and W. D. O'Brien (2012). "Development of a theoretical model describing sonoporation activity of cells exposed to ultrasound in the presence of contrast agents." Journal of the Acoustical Society of America **131**(4): 2723-2729.
10. Inserra, C., G. Regnault, S. Cleve, C. Mauger and A. A. Doinikov (2020). "Acoustic microstreaming produced by nonspherical oscillations of a gas bubble. III. Case of self-interacting modes n-n." Physical Review E **101**(1).
11. Katiyar, A. and K. Sarkar (2012). "Effects of encapsulation damping on the excitation threshold for subharmonic generation from contrast microbubbles." Journal of the Acoustical Society of America **132**(5): 3576-3585.

12. Krasovitski, B. and E. Kimmel (2004). "Shear stress induced by a gas bubble pulsating in an ultrasonic field near a wall." Ieee Transactions on Ultrasonics Ferroelectrics and Frequency Control **51**(8): 973-979.

13. Lentacker, I., S. C. De Smedt and N. N. Sanders (2009). "Drug loaded microbubble design for ultrasound triggered delivery." Soft Matter **5**(11): 2161-2170.

14. Lewin, P. A. and L. Bjo/rno/ (1982). "Acoustically induced shear stresses in the vicinity of microbubbles in tissue." The Journal of the Acoustical Society of America **71**(3): 728-734.

15. Lighthill, J. (1978). "Acoustic Streaming." Journal of Sound and Vibration **61**(3): 391-418.

16. Liu, R. H., J. Yang, M. Z. Pindera, M. Athavale and P. Grodzinski (2002). "Bubble-induced acoustic micromixing." Lab on a Chip **2**(3): 151-157.

17. Marmottant, P. and S. Hilgenfeldt (2003). "Controlled vesicle deformation and lysis by single oscillating bubbles." Nature **423**(6936): 153-156.

18. Marmottant, P., J. Raven, H. Gardeniers, J. Bomer and S. Hilgenfeldt (2006). "Microfluidics with ultrasound-driven bubbles." Journal of Fluid Mechanics **568**: 109-118.

19. Marmottant, P., S. van der Meer, M. Emmer, M. Versluis, N. de Jong, S. Hilgenfeldt and D. Lohse (2005). "A model for large amplitude oscillations of coated bubbles accounting for buckling and rupture." Journal of the Acoustical Society of America **118**(6): 3499-3505.

20. Meng, L., X. Liu, Y. Wang, W. Zhang, W. Zhou, F. Cai, F. Li, J. Wu, L. Xu, L. Niu and H. Zheng (2019). "Biomedical Ultrasound Bioeffects: Sonoporation of Cells by a Parallel Stable Cavitation Microbubble Array (Adv. Sci. 17/2019)." Advanced Science **6**(17): 1970099.

21. Mobadersany, N. and Sarkar K. (2018). "Collapse and Jet Formation of Ultrasound Contrast Microbubbles near a Membrane for Sonoporation." In 10th International Cavitation Symposium, Baltimore, MD, USA. ASME.

22. Mobadersany N, Sarkar K. (2018). "The dynamic of contrast agent and surrounding fluid in the vicinity of a wall for sonoporation." *ArXiv Preprint ArXiv:180208652*.

23. Mobadersany, N., & Sarkar, K. (2019). Acoustic microstreaming near a plane wall due to a pulsating free or coated bubble: Velocity, vorticity and closed streamlines. *Journal of Fluid Mechanics, 875*, 781-806. doi:10.1017/jfm.2019.478

24. Mobadersany, N. and K. Sarkar (2019). "Acoustic microstreaming near a plane wall due to a pulsating free or coated bubble: velocity, vorticity and closed streamlines." Journal of Fluid Mechanics **875**: 781-806.

25. Mobadersany, N., Sarkar, K. (2019). "Encapsulated microbubbles for contrast ultrasound imaging and drug delivery: from pressure dependent subharmonic to collapsing jet and acoustic streaming." 4th Thermal and Fluids Engineering Conference, ASTFE Digital Library, Pages 817-828, Doi: 10.1615/TFEC2019.bio.028506

26. Nyborg, W. L. (1953). "Acoustic Streaming due to Attenuated Plane Waves." The Journal of the Acoustical Society of America **25**(1): 68-75.

27. Nyborg, W. L. (1958). "Acoustic Streaming near a Boundary." The Journal of the Acoustical Society of America **30**(4): 329-339.

28. Orbay, S., A. Ozcelik, J. Lata, M. Kaynak, M. X. Wu and T. J. Huang (2017). "Mixing high-viscosity fluids via acoustically driven bubbles." Journal of Micromechanics and Microengineering **27**(1).

29. Paul, S., A. Katiyar, K. Sarkar, D. Chatterjee, W. T. Shi and F. Forsberg (2010). "Material characterization of the encapsulation of an ultrasound contrast microbubble and its subharmonic response: Strain-softening interfacial elasticity model." Journal of the Acoustical Society of America **127**(6): 3846-3857.

30. Pommella, A., N. J. Brooks, J. M. Seddon and V. Garbin (2015). "Selective flow-induced vesicle rupture to sort by membrane mechanical properties." Scientific Reports **5**.

31. Rallabandi, B., C. Wang and S. Hilgenfeldt (2014). "Two-dimensional streaming flows driven by sessile semicylindrical microbubbles." Journal of Fluid Mechanics **739**: 57-71.

32. Rayleigh, L. (1945). The theory of sound; with a historical introduction by Robert Bruce Lindsay, 2nd ed. Oxford, England, Dover Publications.

33. Riley, N. (2001). "Steady streaming." Annual Review of Fluid Mechanics **33**: 43-65.

Rooney, J. A. (1970). "Hemolysis near an Ultrasonically Pulsating Gas Bubble." Science **169**(3948): 869-&.

34. Roovers, S., T. Segers, G. Lajoinie, J. Deprez, M. Versluis, S. C. De Smedt and I. Lentacker (2019). "The Role of Ultrasound-Driven Microbubble Dynamics in Drug Delivery: From Microbubble Fundamentals to Clinical Translation." Langmuir **35**(31): 10173-10191.

35. Scriven, L. E. (1960). "Dynamics of a fluid interface Equation of motion for Newtonian wall fluids." Chemical Engineering Science **12**(2): 98-108.

36. Shervani-Tabar, M. T. and N. Mobadersany (2013). "Numerical study of the dielectric liquid around an electrical discharge generated vapor bubble in ultrasonic assisted EDM." Ultrasonics **53**(5): 943-955.

37. Shervani-Tabar, M. T., N. Mobadersany, S. M. S. Mahmoudi and A. Rezaee-Barmi (2011). "Velocity field and pressure distribution around a collapsing cavitation bubble during necking and splitting." Journal of Engineering Mathematics **71**(4): 349-366.

38. Sirsi, S. R. and M. A. Borden (2014). "State-of-the-art materials for ultrasound-triggered drug delivery." Advanced Drug Delivery Reviews **72**: 3-14.

39. Stuart, J. T. (1966). "DOUBLE BOUNDARY LAYERS IN OSCILLATORY VISCOUS FLOW." Journal of Fluid Mechanics **24**: 673-&.

40. Taib, B. B. (1985). "Boundary integral method applied to cavitation bubble dynamics." Doctor of Philosophy, University of Wollongong.

41. Tharkar, P., P. Varanasi, W. S. F. Wong, C. T. Jin and W. Chrzanowski (2019). "Nano-Enhanced Drug Delivery and Therapeutic Ultrasound for Cancer Treatment and Beyond." Frontiers in Bioengineering and Biotechnology **7**.

42. Westervelt, P. J. (1953). "The Theory of Steady Rotational Flow Generated by a Sound Field." The Journal of the Acoustical Society of America **25**(1): 60-67.

43. Wu, J., J. P. Ross and J.-F. Chiu (2002). "Reparable sonoporation generated by microstreaming." The Journal of the Acoustical Society of America **111**(3): 1460-1464.

44. Wu, J. R. and G. H. Du (1997). "Streaming generated by a bubble in an ultrasound field." Journal of the Acoustical Society of America **101**(4): 1899-1907.
16